# Exploratory Analysis of Pairwise Interactions in Online Social Networks

Luka Humski, Damir Pintar and Mihaela Vranić

*University of Zagreb, Faculty of Electrical Engineering and Computing Unska 3, 10000 Zagreb, Croatia*

Mihaela Vranić, mihaela.vranic@fer.hr, +385 98 954 68 87

# Exploratory Analysis of Pairwise Interactions in Online Social Networks

In the last few decades sociologists were trying to explain human behaviour by analysing social networks, which requires access to data about interpersonal relationships. This represented a big obstacle in this research field until the emergence of online social networks (OSNs), which vastly facilitated the process of collecting such data. Nowadays, by crawling public profiles on OSNs, it is possible to build a social graph where “friends” on OSN become represented as connected nodes. OSN connection does not necessarily indicate a close real-life relationship, but using OSN interaction records may reveal real-life relationship intensities, a topic which inspired a number of recent researches. Still, published research currently lacks an extensive exploratory analysis of OSN interaction records, i.e. a comprehensive overview of users’ interaction via different ways of OSN interaction. In this paper we provide such an overview by leveraging results of conducted extensive social experiment which managed to collect records for over 3,200 Facebook users interacting with over 1,400,000 of their friends. Our exploratory analysis focuses on extracting population distributions and correlation parameters for 13 interaction parameters, providing valuable insight in online social network interaction for future researches aimed at this field of study.

## Introduction and related work

A social network is a structure composed of nodes and edges which represent people and their relationships, such as family bonds, friendships, etc. Social network analysis (SNA) is a research field which deals with analysing such networks and extracting useful information about people described within, with the analysis being mostly focused on user interactions. There are numerous possible applications: by analysing social networks sociologists and social psychologists are trying to explain how people’s thoughts, feelings and behaviours are influenced by presence of others [1,2];

recommender systems can use it to make customized and novel recommendations [3,4]; corporations are trying to improve relations between employees and their working effect [5–7]; telecoms want to prevent users churn [8–10]; in the educational domain information about connectedness between students may be used to enhance the learning process [11–13], etc.

Modern online social networks (OSNs) such as Facebook or Twitter are widely accepted as platforms for exchanging messages, sharing photos, links and other kinds of information. We can treat these OSNs as applications for social networks management. Due to their nature as digital platforms, information about connectedness and interaction between users is usually stored in a structured fashion and is becoming more accessible than ever, which has vastly facilitated the ability to observe social networks for research purposes. One of the basic methods of gathering OSN information is creating software which uses the OSN’s API to crawl public profiles and construct a social graph based on publicly available "friendship" information contained within [14–16]. In that way it is possible to create a social graph with information whether two users are connected, but usually not the details about the nature or intensity of their real-life relationship. There are however some researches that introduce various models and algorithms which enable calculating friendship intensity and picking out real-life relationships from ego-users’ total OSN friends by considering their interaction on OSN [17–26]. Some papers aim simply to differentiate between strong and weak friendships of the ego-user [17–19], others classify ego-user’s friends in more than just two basic classes [20,21] while some aim to determine the connection strengths between all OSN users and express it in a numerical fashion [20,22–26].

Although OSN interaction records are frequently used as basis for various research purposes, so far a comprehensive exploratory analysis of users’ OSN interaction has not yet been published. Taken this into consideration, we have decided to invest a great effort in collecting a representative real-life OSN interaction dataset, followed by performing an extensive exploratory analysis in order to extract and describe its key properties. As Facebook is arguably the most popular OSN today with over 2 billion active users [27], we decided to focus on this particular social network. We have conducted a comprehensive Facebook social experiment *NajFrend* where we collected records that describe interaction between almost a million and a half pairs of Facebook users. We have then performed an exploratory analysis where we focused on extracting population distributions and correlation matrices for 13 Facebook interaction parameters such as posts, likes, comments, mutual photos etc. (which we will call *interaction parameters* in the following sections). All these parameters were collected and summarized on pairwise levels – e.g. total likes, total comments, etc. between pairs of Facebook friends. The results of this user interaction exploratory analysis based on huge empirical dataset represents the pivotal contribution of this paper.

The paper is organized as follows: in the *Methodology* section we provide details about the conducted social experiment, present the collected dataset and describe in detail the process of extracting population distributions and constructing the correlation matrix; the *Results* section contains tabular and visual results of the exploratory analysis; *Discussion* provides insight and interpretations of gained results; finally, in *Conclusion* we give final remarks on this research.

---

**Methodology**

This section will provide a brief description of the conducted social experiment *NajFrend* and the dataset collected in that experiment, which is a core dataset for our exploratory analysis. Also, we will explain the steps undertaken in the exploratory analysis itself.

### *Social experiment NajFrend and the collected dataset*

*NajFrend* is a comprehensive social experiment held in April and May of 2015. It has involved 3,277 examinees, mostly from Croatia and neighbouring countries. Majority of examinees were between 18 and 30 years old. Close to 80% of examinees were high school and university students. 57.7% of examinees were men and 42.3% were women. This experiment collected a dataset about interactions between 3,277 examinees and over 1,400,000 of their Facebook friends. All examinees gave explicit permission to allow using collected data about their Facebook interaction for this research.

For the following exploratory analysis, we have chosen 13 Facebook interaction parameters to describe user interactions, whose list and explanations can be found in Table 1. Additionally, for each attribute in the table we have included an abbreviation which will be used in the certain following figures with insufficient space for the full attribute names.

*Table 1 Available interaction parameters*

| **Interaction parameter name** | **Abbreviation** | **Description** |
|---|---|---|
| friend_mutual | fm | Number of mutual friends between ego-user and his observed Facebook friend |
| feed_like | fl | Number of observed friend’s "likes" on ego-user's posts |
| feed_comment | fc | Number of observed friend’s comments on posts on the ego-user's timeline |
| feed_addressed | fa | Number of observed friend’s posts on the ego-user's timeline |
| feed_together_in_post | ftp | Number of times when ego-user and his observed Facebook friend are tagged together in posts |
| mutual_photo_published_by_user | mpu | Number of mutual photos of ego-user and his observed Facebook friend published by ego-user |
| mutual_photo_published_by_friend | mpf | Number of mutual photos of ego-user and his observed Facebook friend published by observed friend |
| mutual_photo_published_by_others | mpo | Number of mutual photos of ego-user and his observed Facebook friend published by some other |
| photo_like | pl | Number of "likes" of observed Facebook friend on ego-user's photos |
| photo_comment | pc | Number of comments by observed Facebook friend on ego-user's photos |
| inbox_chat | ic | Number of exchanged private messages between ego-user and his observed Facebook friend (taking into account only last 50 from ego user's total conversations) |
| my_photo_likes | mpl | Number of ego-user's "likes" on observed Facebook friend’s photos |
| my_link_likes | mll | Number of ego-user's "likes" on observed Facebook friend’s links |

### *Exploratory analysis*

Main goal of our exploratory analysis was to analyse behaviour of the collected Facebook interaction parameters. We focused on extracting population distributions for each of the observed 13 interaction parameters and calculating *Pearson’s correlation coefficient* for each pair of interaction parameters. For each distribution we have provided a detailed quantile table and a theoretical distribution which has shown to be the best approximation for an empirical distribution of each interaction parameter. Since most Facebook users interact very little with a large portion of their Facebook friends, our dataset contains a lot of zero values. Taking this into consideration, we have chosen to focus on the best approximative theoretical distribution for the non-zero values and present ratios of zero-values for each interaction parameter. The following candidate distributions were tested for each parameter: *beta*, *gamma*, *inverse gamma, normal, log-normal, skewed normal, geometrical* and *uniform*. For the theoretical distributions which are defined only on interval [0,1] we have first normalized the data according to (1).

$$x_{norm} = \frac{x_{empirical} - x_{min}}{x_{max} - x_{min}} \qquad (1)$$

*Maximum likelihood estimation* (MLE) was used for each listed distribution to find the distribution parameters which show the best fit. Using the *chi-square test* we have decided on the final theoretical distributions with lowest corresponding *chi-square* values.

## Results

### *Analysis of the underlying distributions*

In this section we will present the results of our underlying distributions analysis for each interaction parameter. Detailed quantile tables with over 10,000 records for each interaction parameter empirical distribution are not included in this paper due to obvious size constrains, but can be found at: *r.lukahumski.iz.hr/EAPIOSN/quantiles.csv*. For each interaction parameter we have found out the best approximative theoretical distribution of non-zero values and presented the ratio of zero-values. Figure 1 shows the results of *chi-square tests* (with the number of bins set to 50) for each interaction parameter for different distributions. To show a simple graphical illustration of differences between empirical distributions and the best approximate theoretical distributions we also include a representative probability density function (PDF) of empirical and approximative theoretical distribution for the *friend_mutual* parameter on Figure 2. Theoretical distribution is depicted as a dotted line, while the empirical distribution is shown with a solid line. In Table 2 we list all the interaction parameters, their best approximative theoretical distribution name, parameters for best fit, resulting *chi-square* value and the ratio of zero values. It is important to emphasise that according to the *chi-square test* it is not unequivocally proven for any interaction parameter to be distributed according to a specific theoretical distribution, but highlighted theoretical distributions are the best approximation for observed empirical distributions considering the scope of observed theoretical distributions.

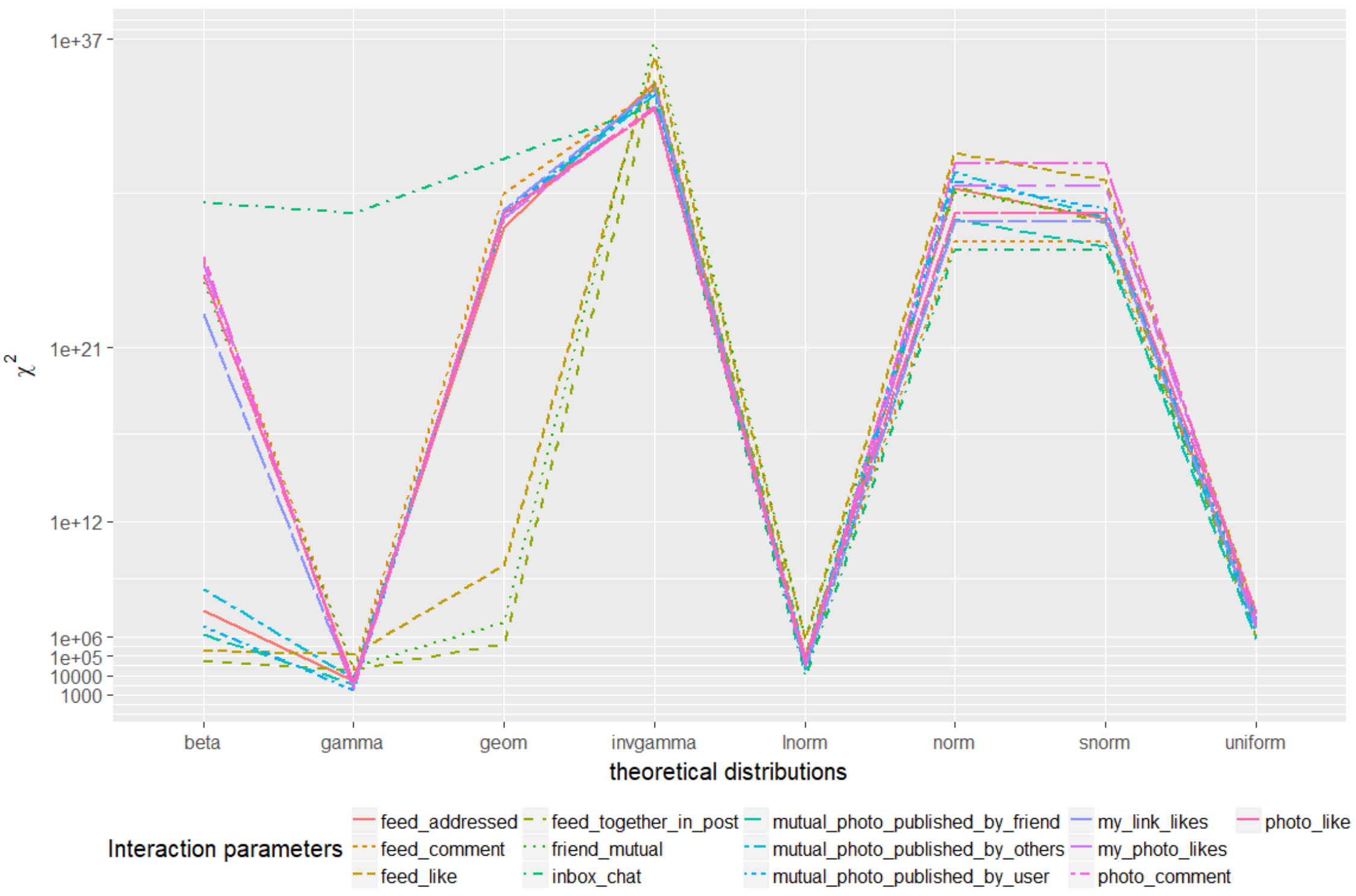

*Figure 1 Results of chi-square test per parameters per distributions*

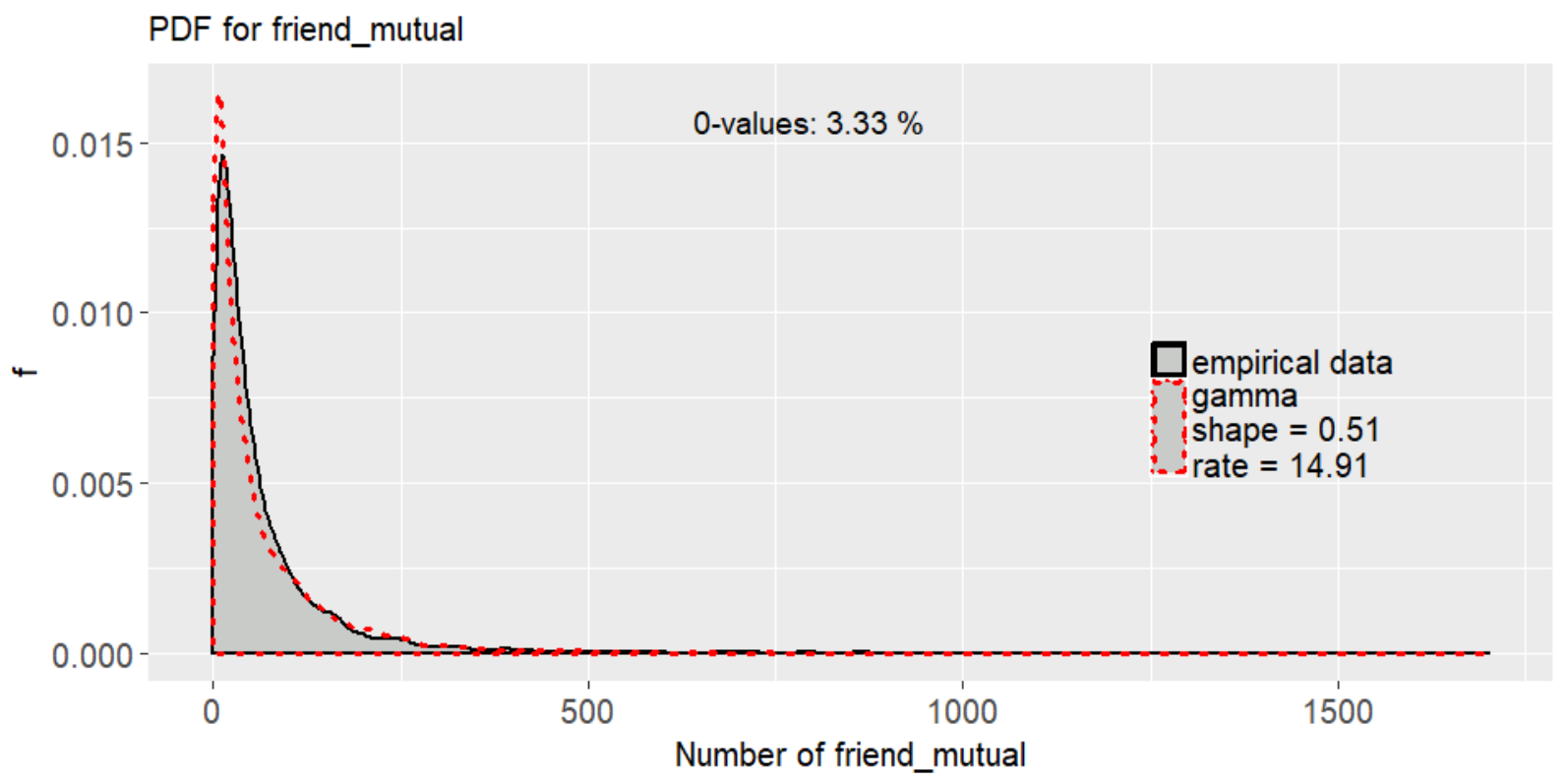

*Figure 2 Comparison of empirical distribution of friend_mutual parameter and the approximative gamma distribution as the best approximative theoretical distribution*

*Table 2 The best approximative theoretical distributions*

| **Interaction parameter name** | **Best approximative theoretical distribution** | **Theoretical distribution parameters** | **Chi-square value** | **Ratio of zero values** |
|---|---|---|---|---|
| friend_mutual | gamma | shape = 0.51<br>rate = 14.91 | 29 369 | 3.33% |
| feed_like | gamma | shape = 0.12<br>rate = 5 | 118 155 | 55.29% |
| feed_comment | gamma | shape = 0.09<br>rate = 17.23 | 17 905 | 86.83% |
| feed_addressed | gamma | shape = 0.07<br>rate = 6.25 | 5 467 | 93.19% |
| feed_together_in_post | gamma | shape = 0.06<br>rate = 4.35 | 20 297 | 97.58% |
| mutual_photo_published_by_user | gamma | shape = 0.07<br>rate = 5.72 | 1 726 | 96.94% |
| mutual_photo_published_by_friend | gamma | shape = 0.07<br>rate = 5.45 | 3 618 | 97.88% |
| mutual_photo_published_by_others | gamma | shape = 0.08<br>rate = 6.98 | 6 851 | 87.35% |
| photo_like | gamma | shape = 0.09<br>rate = 12.07 | 4 359 | 71.67% |
| photo_comment | gamma | shape = 0.09<br>rate = 11.11 | 2 081 | 90.46% |
| inbox_chat | log-normal | meanlog = -8.83<br>sdlog = 4.26 | 9 024 | 91.86% |
| my_photo_likes | gamma | shape = 0.08<br>rate = 14.52 | 4 098 | 82.03% |
| my_link_likes | gamma | shape = 0.07<br>rate = 8.71 | 2 688 | 94.34% |

### *Analysis of correlations between interaction parameters*

*Pearson's correlation coefficients* between attributes in the dataset are shown in Figure 3. Upper part of the figure shows correlation intensity using the size and colour of the squares, while the lower part shows exact numerical values. Due to reasons of clarity all attributes have abbreviated names (according to Table 1).

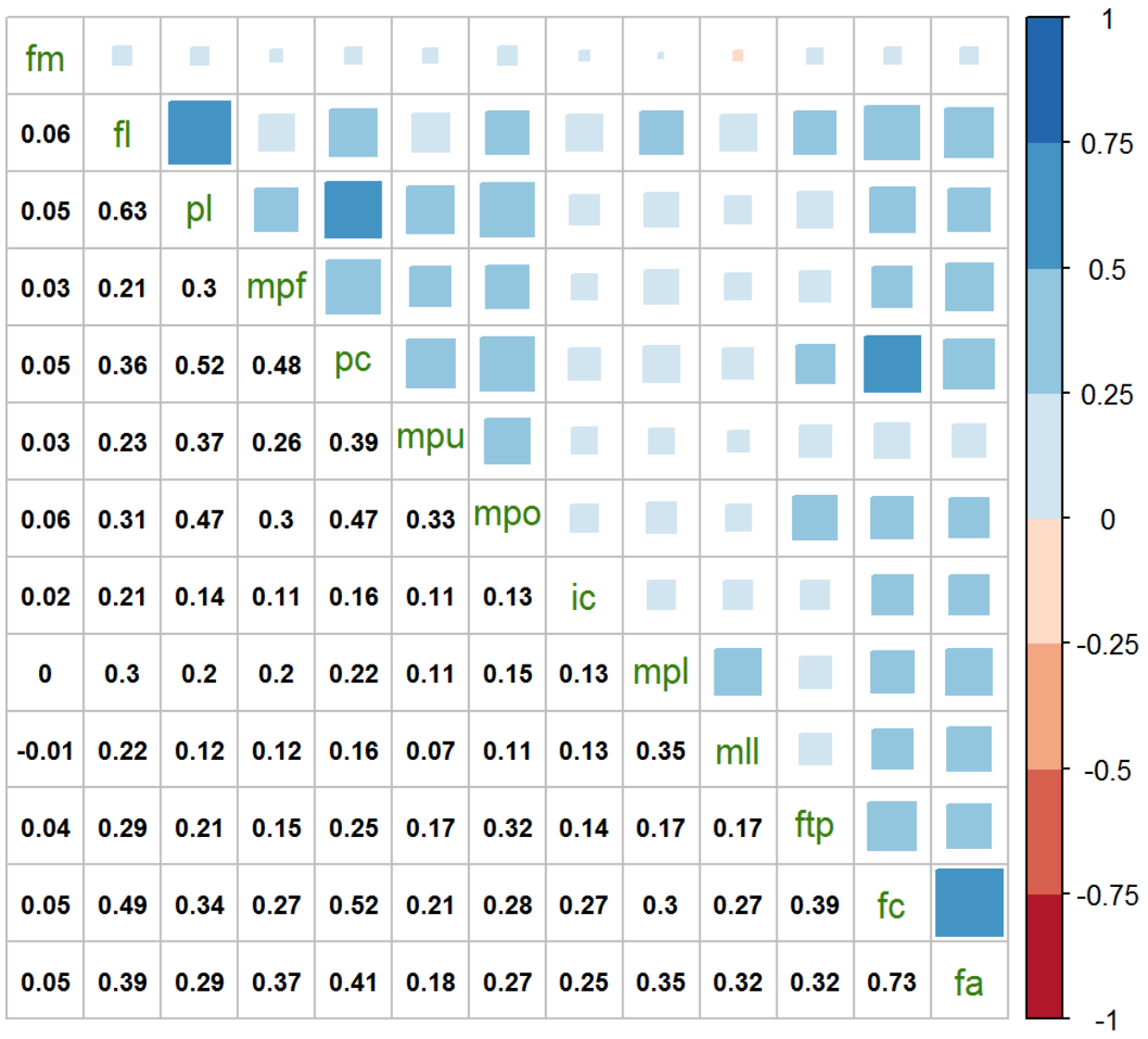

*Figure 3 Correlation between attributes available in dataset*

## Discussion

Previous section presented the results of exploratory analysis done by using the dataset gained in the conducted social experiment. In the following paragraphs, we will briefly review gained results and try to provide some interpretations.

Correlations show which interaction parameters are connected and how strong that connection is. Our analysis shows that *feed_comment* and *feed_addressed* have the strongest correlation. It is interesting to note that people who make a lot of comments on friend’s posts will also write many standalone posts on their respective timelines. Analysis also shows high correlation between parameters *photo_like* and *feed_like*, which is logical concerning the nature of these parameters, i.e. users treat reacting to textual posts and pictures very similarly. High correlation between attributes *photo_comment* and *feed_comment* also supports this assumption.

Low correlation between parameters that show the numbers of mutual photos is slightly surprising. We previously expected to see a relative similarity between parameters *mutual_photo_published_by_user*, *mutual_photo_published_by_friend* and *mutual_photo_published_by_others* because all these parameters count the number of mutual photos between ego-users and their observed Facebook friend, with the only difference being the person who published the photo. Analysis, however, showed that photo sharing habits vary significantly between users.

Another interesting find is that there is no correlation between number of mutual friends and the level of interaction on OSN via observed interaction parameters. An assumption can be made that people who have more friends in common belong to a certain clique which will be reflected in a more intensive on-line communication, but our analysis showed this is not corroborated by facts gained by the survey results.

When looking at various distributions, the large number of zero values is apparent, meaning that ego-users generally interact very little with most of their Facebook friends. This is not so surprising if we refer to the *Dunbar’s number* [28] which states that people can comfortably maintain only 150 stable relationships, compared to the average number of Facebook "friends" in our survey which was 429.

The total lack of interaction further affirms this supposition, and this fact additionally motivates researches which aim to distinguish OSN friends which truly are digital representations of actual real-life relationships.

Finally, if one wants to model interaction paramater behavior using theoretical distributions, the overall best approximative theoretical distribution for all interaction parameters is the *gamma distribution*, the sole exception being the *inbox_chat* parameter for which the *log-normal* distribution gives the best results.

## Conclusion

In this paper we have presented the results of our exploratory analysis aimed to extract key properties of the data which describes interactions between pairs of connected Facebook users. For each interaction parameter we have provided an empirical distribution as a detailed quantile table. Also, we discovered the best approximative theoretical distributions and associated parameters for all observed interaction parameters. For all pairs of interaction parameters, we presented the level of correlation by calculating the *Pearson's correlation coefficient*.

The presented dataset was obtained in a massive social experiment *NajFrend* which involved over 3,000 participants and collected more than 1,400,000 records with summarized frequencies of interaction parameters between ego-users and their Facebook friends. The interaction records were collected using of Facebook API 1.0. This dataset will also be the mainstay of our future research involving methods for discovering and visualizing real-life relationships based on observed social network interaction parameters.

**Acknowledgment**

The research team would like to thank Croatian Science Foundation (*Hrvatska zaklada za znanost* – www.hrzz.hr). The work has been fully supported by Croatian Science Foundation under the project UIP-2014-09-2051 eduMINE – Leveraging data mining methods and open technologies for enhancement of the e-learning infrastructure. We would also like to thank to our ex-student Juraj Ilić who developed PHP application "NajFrend" for conducting survey on which this research is based and helped us in survey conducting.